\documentclass{article}
\usepackage{amsfonts}
\usepackage{graphicx}
\usepackage{epsfig}
\usepackage{eepic}
\usepackage{cite}
\usepackage{graphicx}
\usepackage{dcolumn}
\usepackage{bm}
\usepackage{times,mathptm}

\newcommand{\RR}{\rangle}
\newcommand{\LL}{\langle}

\newcommand{\nc}{\newcommand}
\nc{\rnc}{\renewcommand}
\nc{\ket}[1]{\left | \, #1 \right \rangle}
\nc{\bra}[1]{\left \langle #1 \, \right |}
\nc{\proj}[1]{\ket{#1}\bra{#1}}
\rnc{\vec}{\mathbf}
\nc{\ua}{\uparrow}
\nc{\da}{\downarrow}

\nc{\braket}[2]{\langle\, #1\,|\,#2\,\rangle}
\nc{\half}{\frac{1}{2}}

\nc{\prj}{\mathcal{P}}
\nc{\hilb}{\mathcal{H}}
\nc{\pth}{\mathcal{C}}
\nc{\inprod}[2]{\braket{#1}{#2}}
\nc{\upket}{\ket{\uparrow}}
\nc{\downket}{\ket{\downarrow}}
\nc{\upbra}{\bra{\uparrow}}
\nc{\downbra}{\bra{\downarrow}}

\begin{document}

\title{Scalable quantum computation in systems with Bose-Hubbard dynamics}
\author{
Guido Pupillo \and
Ana Maria Rey \and
Gavin Brennen \and
Carl J. Williams \and
Charles W. Clark 
\\
National Institutes of Standards and Technology, \\
Gaithersburg, MD, 20899}
\maketitle

\begin{abstract}
Several proposals for quantum computation utilize a lattice type architecture with qubits trapped by a periodic potential.  For systems undergoing many body interactions described by the Bose-Hubbard Hamiltonian, the ground state of the system carries number fluctuations that scale with the number of qubits.  This process degrades the initialization of the quantum computer register and can introduce errors during error correction.  In an earlier manuscript we proposed a solution to this problem tailored to the loading of cold atoms into an optical lattice via the Mott Insulator phase transition.  It was shown that by adding an inhomogeneity to the lattice and performing a continuous measurement, the unit filled state suitable for a quantum computer register can be maintained.  Here, we give a more rigorous derivation of the register fidelity in homogeneous and inhomogeneous lattices and provide evidence that the protocol is effective in the finite temperature regime.
\end{abstract}

\section{Introduction} \label{sIntro}
The goal of building a quantum computer has spurred tremendous progress in coherent control and measurement of small quantum systems.  In order to fully realize the promised computational speedup of a quantum device, the underlying system should be scalable to a large number of information carriers or qubits.  Indeed, the first two criteria delineated by DiVincenzo \cite{DiVincenzo:00} for scalable quantum computation are:
\begin{itemize}
\item {A scalable physical system with well characterized qubits}  
\item {The ability to initialize the state of the qubits to a simple fiducial state}
\end{itemize}
In many systems, the first criterion can be met by increasing the number of storage components for the qubits, e.g. in solid state systems the number of dopant qubits in the bulk material could be increased, or for ions large scale micro-trap arrays have been proposed \cite{Wineland:02}.  There are two main approaches to satisfy the second criterion \cite{DiVincenzo:00}.  One is to allow the system to interact with the environment and ``naturally" cool to its ground state and thereafter use this state as the initial state.  The other is to actively cool each qubit by projective measurement to a fiducial state $|0\rangle$.  A problem arises with these approaches when the ground state of the many body Hamiltonian is not a suitable initial state.  This is the case for bosonic qubits embedded in systems with periodic confinement, for example in josephson junction arrays \cite{van} and neutral atoms trapped in electromagnetic microtraps \cite{Kruger:03} or optical lattices \cite{Greiner}.  For these systems, the underlying dynamics is Bose-Hubbard like and the ground state contains residual coherences described by non-zero number fluctuations in each mode.  We show in Sec. \ref{sBose} that these fluctuations scale with the number of qubits so that if not corrected, the dynamics will impart a constraint to scalability.  In many cases, the tunneling energy between lattice sites is weak so that the overall error probability per qubit is small, however, if the system is projected onto a local number state basis during error correction cycles, the error can become appreciable \cite{Knill}.

We propose a solution to this problem which reduces number fluctuations to negligible size even for large numbers of qubits.  The technique is exemplified in a three dimensional optical lattice filled with cold neutral bosonic atoms.  Optical lattices are three dimensional potentials created by counter-propagating beams of laser light that trap atoms by the conservative electric dipole potential \cite{Deutsch:96}.  This is a particularly attractive system to study because it has been theoretically shown \cite{Jaksch}, and recently demonstrated in the laboratory \cite{Greiner}, that an optical lattice can be loaded from an atomic BEC.  If  the lattice is turned on adiabatically with respect to low lying many body excitations, then the superfluid-like BEC undergoes a phase transition to the Mott insulator (MI) state characterized by predominantly the same number of atoms in each lattice well \cite{Fisher}.  A difficulty with using this mechanism to prepare a unit filled register of atomic qubits arises due to imperfect filling of the lattice as well as number fluctuations intrinsic to the Bose-Hubbard dynamics.

In \cite{BPRCW} we described how to eliminate number fluctuations in the many body ground state.  The approach was to first introduce a quadratic magnetic trap that acts to fill sites with missing qubits in the center, and second to project out components of the many-body wavefunction with multiply occupied lattice sites.  In this paper we give a more rigorous derivation of the dynamics due to first order couplings outside the unit filled state in both homogeneous and inhomogeneous systems.  The essential result is that deep in the Mott Insulator regime, the dynamics can be effectively modeled with a restricted basis involving two level couplings between the unit filled state and first excited states with double occupancy in one lattice site.  We then show in Sec. \ref{sMeas} that by turning on an external field components of the many body state with multiply occupied wells can be projected out.  Provided the measurement strength is sufficiently large, the system does not evolve out of the restricted basis and the measurement can maintain a unit filled register for the duration of quantum computation.  Finally, we give evidence that the proposal can be made robust even for finite temperature systems.                 

\section{Bose-Hubbard Dynamics}
\label{sBose}

\noindent

The Bose-Hubbard Hamiltonian describes a system of bosons on a lattice having a kinetic energy associated with tunneling between nearest neighbor lattice sites and a pairwise interaction at each lattice site: 

\bigskip

\begin{equation}
H_{BH}=-\sum_{<\mathbf{i,j}>}J(\mathbf{j})a_{\mathbf{j}}^{\dagger }a_{\mathbf{i}}+\sum_{%
\mathbf{j}}\epsilon (\mathbf{j})n_{\mathbf{j}}+\frac{U(\mathbf{j})}{2}n_{\mathbf{j}}(n_{%
\mathbf{j}}-1)\;
\end{equation}
Here $a_{\mathbf{j}}$ are the bosonic annihilation operators and $n_{\mathbf{%
j}}=a_{\mathbf{j}}^{\dagger }a_{\mathbf{j}}$ the number operators for an
atom in the lowest vibrational state of lattice well $\mathbf{j=(}
j_{1},j_{2,...},j_{d})$ with $d$ the dimensionality of the \ lattice.  The
energy offset at each lattice site is $\epsilon (\mathbf{j})$ which models a
continuously varying external potential.  Henceforth we assume that 
the energies $J$ and $U$ describing
tunneling and on-site interaction energies respectively are site-independent.
The notation
$<\mathbf{i,j}>$ indicates that the sum is over nearest neighbors.  In the
tight-binding model, the nearest neighbor tunneling energy $J$ is defined as
one fourth the band width of the lowest occupied band.  The dynamics of an 
ultracold  bosonic gas in an optical lattice
loaded in such a way that only the lowest vibrational state is
occupied can be described by $H_{BH}$ where the
system parameters are controlled by laser light.
In a real optical lattice,
tunneling in one, two, or three dimensions can be achieved by tuning the laser intensities and
detunings so that dynamics can be effectively frozen in some directions.  We assume a square lattice 
with tunneling 
dynamics in $d$ dimensions through a separable potential barrier 
$V(\mathbf{x})=V\sum_{i=1}^{d}\cos ^{2}(kx_{i})$.  The tunneling rate is closely
approximated by $J/\hbar =4/(\sqrt{\pi }\hbar )E_{R}(V/E_{R})^{3/4}e^{-2\sqrt{V/E_{R}}}$
 \cite{MF}, where the recoil energy is $E_{R}=(\hbar k)^{2}/2m$ ($m=$atomic mass). The
on site interaction is a result of the ground state collisions described by
the $s$-wave scattering length $a_{s}$ between two atoms each in the motional
state $\phi (\mathbf{x})$ and is given by $U=\frac{4\pi a_{s}\hbar ^{2}}{m}
\int d\mathbf{x}|\phi (\mathbf{x})|^{4}$.

It was recognized early on that the MI transition might be an efficient way
to initialize a register of atomic qubits in an optical lattice for use in
quantum information processing. A key advantage of loading from a BEC is the
availability of an initially high phase space density which can be frozen to
the MI state with atoms occupying most lattice sites. For the homogeneous
system ($\epsilon (\mathbf{j})=0)$, only commensurate fillings give rise to
a MI transition.  For the purposes of quantum computation, one particle per well is 
desirable.  In practice, this is difficult to achieve directly because the precise number of
atoms is unknown and the lattice strength is not perfectly uniform on the boundaries.
There are proposals to prepare unit filled lattices using dissipative techniques involving 
filling the lattice atom by atom \cite{Weiss} or using Raman side-band cooling \cite{Jessen}.
Additionally, it has been shown that one can repair imperfect filling from a BEC via an adiabatic transfer mechanism between two sublevels of each atom \cite{Zoller}.  While these techniques can efficiently initialize the
lattice, the unit filled register is not a stationary state of the system and the dynamics resulting from
the inter-well coupling degrades the register fidelity defined as the population in the unit filled state.  In Sec. \ref{sHomap} we discuss the dynamics in a homogeneous system assuming that an initial state with unit filling has been prepared. We show that in the strong coupling limit, $J/U\ll 1$, the dynamics can be restricted to a subspace whose dimension grows linearly with the number of particles in the system.  The fidelity, in this limit, is then derived explicitly. In Sec. \ref{sDyn} we show that an external quadratic potential 
can be used to help prepare a unit filled state in an appropriate subspace of the lattice, which will constitute a quantum computer register. The dynamical fidelity derived in the homogeneous case can then be mapped to the one in the register.

\subsection{ Homogeneous approximation}
\label{sHomap}

In order to understand the relevant dynamics in our system we first discuss the properties 
of an idealized homogeneous lattice with periodic boundaries and unit filling, $N=M$ where 
$N$ is the number of atoms and $M$ is the number of wells.
The regime of interest is the strong coupling limit, 
which is experimentally achievable in an optical lattice because the tunneling decreases
exponentially with the trap depth.  To zero order, neglecting the kinetic
energy, the ground state of the system is the unit filled
state $\left| T\right\rangle=\prod_{{\mathbf j}}a_{{\mathbf j}}^{\dagger}|0\rangle $ which is the Fock state with exactly
one atom per well and energy zero.  A small but non-zero $J/U$ introduces number fluctuations in the ground state 
which provide a small residual
coherence across the system \cite{Roberts,Guido}.  We can calculate the approximate ground state by including states which couple to
$|T\rangle$ to second order in $H_{BH}$.  The first  $N(N-1)$ degenerate excited states are
Fock states with two particles in one well, one hole in another well and unit filling 
in every other site.  We define such states particle-hole excitations and their energy is equal to $U$.
To first order, $|T\rangle$ couples only to a state with a doubly occupied well neighboring the hole.  Because $H_{BH}$ is invariant under translations of the lattice, this state, when properly symmetrized, is
\begin{equation}
\left| S_{1}\right\rangle\equiv\sum_{<{\mathbf i},{\mathbf j}>}\frac{a_{{\mathbf i}}^{\dagger}a_{{\mathbf j}}|T\rangle }{\sqrt{4dN}}\;
\end{equation}
To second order, the coupling is to three symmetrized states:   the state with two particle-hole excitations $|C_{11}\rangle$, the state with three particles in one site and holes in two neighboring sites $|Q_1\rangle$, and $|S_2\rangle$, the symmetrized state where the doubly occupied site and the hole are next nearest neighbors.  The former two states are given by\\
$|C_{11} \rangle \equiv f(N,d)\sum_{ < {\mathbf i}, {\mathbf j}>\neq < {\mathbf p}, {\mathbf q}>}
a^{ \dagger }_{{\mathbf i}} a_{ {\mathbf j}}a^{ \dagger }_{{\mathbf q}} a_{ {\mathbf p}} |T\rangle$, 
$|Q_{1} \rangle \equiv g(N,d) \sum_{ < \mathbf{i j k} > } a_{{\mathbf i}} a^{ \dagger 2}_{{\mathbf j}} a_{{\mathbf k}}  | T \rangle$,
and have energies $2U$ and $3U$ respectively.  The second order coupling is proportional to 
the inverse of the normalization factors of the three states:  for $|C_{11}\rangle$, the normalization is $f(N,d)=(2dN(dN-3-4(d-1)))^{-1/2}$, for $|Q_1\rangle$ the normalization is $g(N,d)= (6d (2d-1) N)^{-1/2}$, and for $|S_2\rangle$ it is $(4dN)^{-1/2}$.  Provided $dN\ll (U/J)^2$, then the perturbation expansion is convergent and to first order the ground state is approximately,
\begin{eqnarray}
\left| \Psi _{g}\right\rangle  &=&\alpha \left( \left| T\right\rangle
+2\frac{J}{U}\sqrt{dN}\left| S_{1}\right\rangle \right)\;
\end{eqnarray}
The normalization constant is $\alpha
=(1+4Nd(J/U)^{2})^{-1/2}.$ To second order, the ground state energy is $E=-4dNJ^{2}/U$.  Notice the fidelity $F=|\langle T|\psi_g\rangle|^2=\alpha^2$ decreases linearly with the number of atoms for fixed $J/U$.

\subsection{Homogenous dynamics}
Any mechanism used to prepare a register of qubits in the unit filled state will suffer a degradation in fidelity, because the state $|T\rangle$ is not a stationary state of the system.  In this section we derive the time dependent fidelity in a homogeneous commensurately filled system prepared at time $t=0$ in the state $|T\rangle$.  While our results can be generalized to higher dimensions,
in the remainder of this section we assume tunneling along one dimension only.  To compute the first order coupling outside of $|T\rangle$ we need to rediagonalize $H_{BH}$ in the subspace spanned
by the states with one particle-hole pair. In principle we
should consider all $N(N-1)$ states. However by translational invariance of $H_{BH}$, 
the dynamics is restricted to the $\lfloor N/2\rfloor$ invariant states in the subspace, where 
$\lfloor x \rfloor$ denotes the greatest integer less than or equal to $x$. 
The states in the particle-hole subspace are represented as
\begin{equation}
\left| \Psi _{r}\right\rangle=\sum_{n=1}^{\lfloor N/2\rfloor
}s_{n}^{r}\left| S_{n}\right\rangle   \label{sim},\ \  
\left| S_{n}\right\rangle\equiv\sum_{|i-j|=n}\frac{a^{\dagger}_ia_j|T\rangle }{\sqrt{4N}}.\;
\end{equation}
In order to diagonalize the Hamiltonian
the coefficients $s_{n}^{r}$ must satisfy the following recurrence relation
\begin{equation}
-3J(s_{n+1}^{r}+s_{n-1}^{r})=(E_{r}-U)s_{n}^{r};\ \ \  s_{0}^{r} =0,s_{i\pm N}^{r}=s_{n}^{r}.\;
\end{equation}
The solutions of the above equations can be shown to be given by
\begin{eqnarray}
s_{n}^{r} &=&\frac{2}{\sqrt{N}}\left| \sin \left( \frac{\pi (2r-1)}{N}
n\right) \right| ,\;r\;=1,..,\lfloor N/2\rfloor  \\
E_{r} &=&U-6J\cos \left( \frac{\pi (2r-1)}{N}\right)
,\;r\;=1,..,\lfloor N/2\rfloor.
\end{eqnarray}
The state vector
at any time can be written as:
\begin{equation}
\left| \Psi (t )\right\rangle=c_{0}(t )\left| T\right\rangle
+\sum_{r=1}^{\lfloor N/2\rfloor }c_{r}(t )\left| \Psi
_{r}\right\rangle\ \ \mathrm{with}\ \  c_{0}(0)=1.\;
\end{equation}

Solving the Schr\"odinger equation to first order in perturbation theory, 
the time dependent fidelity \ to be
in the unit filled state $F(t)=|c_{0}(t )|^{2}$ can be estimated to be
\begin{eqnarray}
F(t) &=&1-\sum_{r=0}^{\lfloor N/2\rfloor}|c_{r}(t )|^{2} \label{sol} \\
|c_{r}(t)|^{2} &\approx &\frac{16}{N}\left( \frac{J}{E_{r}}
\right) ^{2}\sin ^{2}\left( \frac{\pi (2r+1)}{N}\right) \sin ^{2}\left(
\frac{E_{r}t/\hbar}{2}\right)   
\end{eqnarray}
The sum in Eq. \ref{sol} can be solved in the strong coupling regime
of interest and the closed form expression is
\begin{equation}
F(t)=1-8\left( \frac{J}{U}\right) ^{2}N\left( 1-\frac{\cos
(Ut/\hbar )J_{1}(6tJ/\hbar )}{3Jt/\hbar }\right),
\label{ana}
\end{equation}
where $J_{n}$ is the $n$th Bessel function of first kind. This solution
compares well with exact numerical simulations for small number of
wells $(N<8)$ and for larger $N$ in the subspace with at most one
site with two atoms. The behavior of $F(t)$ is determined by fast
oscillations with frequency equal to one, modulated by longer
oscillations with frequency determined by the zeros of $J_{1}$.
For short times, $Jt/\hbar \ll 1$, the fidelity is
\begin{equation}
F(t)\approx 1-16\left( \frac{J}{U}\right) ^{2}N\sin ^{2}\left( \frac{Ut/\hbar }{2}
\right),\;
\label{shortfid}
\end{equation}
which corresponds to the Rabi oscillations of an effective two level system
spanned by $\left| T\right\rangle $ and $\left| S_{1}\right\rangle .$ \
For later times the coupling to the other states become important.  Notice that 
fidelity when time averaged over many oscillation periods $2\pi\hbar/U$ is $\LL F(t)\RR=1-8(J/U)^2N$.
Consequently, the deviation from a perfectly prepared register, described by $1-\LL F(t)\RR$, is twice as bad 
as if the system were prepared in the stationary ground state $|\Psi_g\rangle$ of the Bose-Hubbard Hamiltonian.  This indicates that a lattice filled with a 
commensurate number of atoms in the Mott-Insulator state may create a more robust quantum 
computer register than one prepared dissipatively.
\subsection{Dynamics in presence of the external trap}
\label{sDyn}
\begin{figure}
\begin{center}
\includegraphics[width=0.75\linewidth]{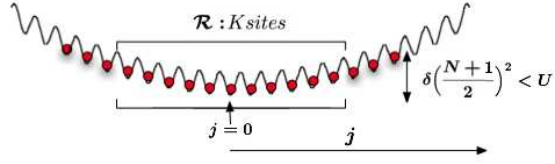}
\caption{\label{fig1}Schematic of an inhomogeneous lattice filled with $N$ qubits with an onsite interaction energy $U$.  An externally applied trapping potential of strength $\epsilon(j)=\delta j^2$, e.g. due to a magnetic field, acts to fill gaps in the central region of the trap.  The center subspace $\mathcal{R}$ of the lattice defines the quantum computer register containing $K<N$ qubits.}
\end{center}
\end{figure}

As mentioned in Sec. \ref{sBose} in practice it is difficult to prepare an optical lattice 
with a commensurate number of atoms and wells.  It can be arranged so that $N<M$ in which case 
we propose to use an inhomogeneous lattice with open boundaries created by
a weak quadratic magnetic trap.  The addition of the trap acts to collect atoms near the potential minimum and leaves empty wells (holes) at the edges.  In experiments where the optical lattice is
loaded from a BEC, the external trap is already present to confine the condensate.  For simplicity we assume a
spherical trap with oscillation frequency in each dimension given by $\omega
_{T}$. The magnetic confinement introduces a characteristic energy scale
$\delta =m/2(\pi /k)^{2}\omega _{T}^2$ so that $\varepsilon (\mathbf{j})=\delta
\sum_{i=1}^{d}j_{i}^{2}$.  In order to insure that the on site interaction energy $U$ 
is larger than the trapping energy of the most externally trapped atoms, we require that the 
trap strength satisfy
$U>\delta (d/2)!N^{2/d}/\pi^{d/2}$. 
Multiple atom occupation is therefore inhibited in any well 
in the ground state configuration. The register is defined by the 
subspace $\mathcal{R} $ comprising \ $K<N$ wells in the
center region of the the trap (see Fig.~\ref{fig1}).  The barrier space flanking $\mathcal{R} $ will
act to suppress percolation of holes from the edges to the center.  The
probability for holes in $\mathcal{R}$ due to tunneling through the
barrier is estimated by computing the product of transition probabilities arising from first order couplings from the boundary of the register at radius $r_{min}$ to the outside edge of the occupied lattice at radius $r_{max}$.   Accounting for tunneling from any site at the surface of the hyper-spherical shell at $r_{max}$, we find $p_{h}\sim \prod_{|\mathbf{j}|=r_{min}}^{r_{max}}dr_{max}^{d-1}\pi^{d/2}J^{2}(\delta
(2|\mathbf{j}|+1))^{-2}/(d/2)!=[(\Gamma[r_{min}+1]/\Gamma[r_{max}+2])^2dr_{max}^{d-1}\pi^{d/2}/(d/2)!][J/(2\delta)]^{(2r_{max}-2r_{min}+2)}$.  The percolation probability $p_h$
is negligible provided the barrier region is
sufficiently large and $J/(2r_{min}\delta)<1$.

Hereafter, we restrict our attention to the dynamics of the reduced state of the register:  $\rho(t)=\mbox{Tr}_{j \notin \mathcal{R}}[\sigma(t)]$ obtained by the tracing over spatial modes outside the register subspace of the entire many body wavefunction $\sigma$.  As with the analysis of dynamics in the homogeneous case, for simplicity we consider tunneling in one dimension only.  The degree of inhomogeneity is 
quantified by the ratio $\delta/J$.  For $0<\delta/J<1$, the energy splitting between Fock states describing
particle-hole pairs is small and the model of homogeneous dynamics is valid.
For $\delta/J>1$, to first order in perturbation theory, the state space is spanned by the unit filled 
state in the register subspace, $|T\rangle_{\mathcal{R}}$, and the $2K$ nearest neighbor particle-hole pairs $|S_{j}^{\pm}\rangle$:
\begin{eqnarray}
|T\rangle_{\mathcal{R}} =\prod_{j=-(K-1)/2}^{(K-1)/2}a_{j}^{\dagger}|0\rangle,\ \ 
|S_{j}^+\rangle_{\mathcal{R}}=\frac{a_{j}^{\dagger}a_{j+1}}{\sqrt{2}}|T\rangle_{\mathcal{R}},\ \ 
|S_{j}^-\rangle_{\mathcal{R}}=\frac{a_{j+1}^{\dagger}a_{j}}{\sqrt{2}}|T\rangle_{\mathcal{R}}.\;
\label{basis}
\end{eqnarray}
Here we have introduced coordinates with the site $j=0$ 
coincident with the trap minimum.
For each $j$ the states $|S_j^{\pm}\rangle_{\mathcal{R}}$ are 
distinguished by the two energetically distinct orientations of a 
doubly occupied site and its neighboring hole with energies, 
$E(S_j^{\pm})=U(1\mp \frac{\delta}{U} (2 j-1))$.  
We define the zero of energy coincident with the state $|T\rangle$. 

We proceed to estimate the fidelity to be in the unit filled state in the register:  
$F_{\mathcal{R}}(t)=\mbox{Tr}[\sigma(t) |T\RR_{\mathcal{R}}{_{\mathcal{R}}\LL}T|]$.  With an eye to the projective measurement, it is important 
to understand the free dynamics when the trap is present and the register is initialized in the pure state, 
$\rho(0)=|T\rangle_{\mathcal{R}}{_{\mathcal{R}}\LL}T|$.  The quantity $F_{\mathcal{R}}(t)$ is generally difficult to compute because atoms couple into and out of the register.  We can solve the simpler problem of the fidelity to be in the target state of a commensurately filled inhomogeneous lattice.  This corresponds to a situation where the register occupies the full extent of a lattice $(K=N)$ with periodic boundary conditions.  In this case, to first order in $J/U$, the time dependent state can be written:
\begin{eqnarray}
|\Psi(t)\rangle=c_T(t)|T\rangle_{\mathcal{R}}+\sum_{j,\pm}e^{-iE(S_j^{\pm})t/\hbar}c_{S_j^{\pm}}(t)|S_j^{\pm}\rangle_{\mathcal{R}}.\;
\end{eqnarray}
The fidelity $F_{com}(\delta,K,t)\equiv |\langle T|\Psi(t)\rangle|^2$ can then be solved as in the homogeneous case.  For $\delta\ll U$ and assuming the $N$-body state is prepared in $|T\rangle$ it is given by
\begin{eqnarray}
F_{com}(N,t)=1-8(J/U)^2(N-\cos(Ut/\hbar)(1+\sin(\delta (N-1)t/\hbar)/\sin(\delta t/\hbar))).\;
\label{fidelity}
\end{eqnarray}
\noindent
This solution compares well with exact numerical simulations 
for a small number of atoms $(N<8)$ and simulations for 
larger commensurate lattices with infinitely high boundaries 
using a restricted basis set of dimension $N(N-1)+1$ consisting of the target state and all particle-hole pairs.  Notice that for $\delta t\rightarrow 0$, 
we recover the short time fidelity in the homogeneous case, Eq. \ref{shortfid}. 
In the incommensurate case we can bound the fidelity $F_{\mathcal{R}}(t)$ inside the register.  Provided $N>K$ the following inequalities on the time averaged fidelities hold:
\begin{equation}
\LL F_{com}(N,t|\sigma(0)=|T\RR\LL T|)\RR \leq \LL F_{\mathcal{R}}(t|\rho(0)=|T\RR_{\mathcal{R}}{_{\mathcal{R}}\LL} T|)\RR\leq \LL F_{com}(K,t|\sigma(0)=|T\RR\LL T|)\RR.\;
\end{equation}
The lower bound arises because the probability to be in the unit filled state of a large commensurately filled lattice, given an initial state which is unit filled, is always less than or equal to the probability to be unit filled over a smaller subspace of $K<N$ of a non commensurately filled lattice whose register is prepared in the unit filled state.  This inequality holds provided the probability for holes to tunnel into the register $\mathcal{R}$ is small over relevant time scales.  The upper bound is a consequence of the fact that unit filling in the register state is degraded because particles can tunnel in and out of the register.  Therefore its fidelity is less than that of a commensurately filled lattice of the same size $K$ prepared in the unit filled state.
\section{Measurement}
\label{sMeas}
In Sec. \ref{sDyn} we showed that the system dynamics can be restricted to a set of two level couplings $\{|T\rangle_{\mathcal{R}}\rightarrow |S_j^{\pm}\rangle\}$.  We now sketch how to perform a continuous measurement to drive the register into the unit filled ``target" state $|T\rangle_{\mathcal{R}}$.  The full details are contained in \cite{BPRCW}.  The idea is to apply an external control field that is resonant with a coupling between the ``faulty" states $\{ |S_j^{\pm}\rangle\}$ and a set of excited states $\{|M_j^{\pm}\rangle_{\mathcal{R}}\}$.  If population in the excited states is easily measurable, for instance by the emission of photons during decay, then the presence of population in the particle-hole states can be monitored.  It is vital that the coupling field be able to spectroscopically resolve the measurement transition without exciting the target state.  This is possible if the atoms in multiply occupied wells see a shifted excited state,  $E(M_j^{\pm})=E(S_j^{\pm})+E-U$, where for example $E$ is  the energy of a dipole-dipole molecular state. The free energy of the system is  
\begin{equation}
\begin{array}{lll}    
H_0+H_{BH}
&=&\sum_{j,\pm}E(S_{j}^{\pm})|S_{j}^{\pm}\rangle_{\mathcal{R}}{_{\mathcal{R}}\langle} S_{j}^{\pm}|+E(M_j^{\pm})|M_j^{\pm}\RR_{\mathcal{R}}{_{\mathcal{R}}\LL} M_j^{\pm}|\\
& &-\sqrt{2}J\sum_{j,\pm}(|S_j^{\pm}\RR_{\mathcal{R}}{_{\mathcal{R}}\LL} T|+|T\RR_{\mathcal{R}}{_{\mathcal{R}}\LL} S_j^{\pm}|).\;
\end{array}
\end{equation}
When the control Hamiltonian with Rabi frequency $\Omega$ is turned on resonant with the measurement transition, the interaction Hamiltonian in the rotating frame is:
\begin{equation}
\begin{array}{lll}
H_I&=&\sum_{j,\pm}((|V_c|+E(S_j^{\pm}))|S_j^{\pm}\rangle_{\mathcal{R}}{_{\mathcal{R}}\langle} S_j^{\pm}|)+(|V_{c}|+E(S_j^{\pm})-U)|M_j^{\pm}\rangle_{\mathcal{R}}{_{\mathcal{R}}\langle} M_j^{\pm}|\\
& &-\sqrt{2}J(|S_j^{\pm}\rangle_{\mathcal{R}}{_{\mathcal{R}}\langle} T|+|T\rangle_{\mathcal{R}}{_{\mathcal{R}}\langle}S_j^{\pm}|)+\hbar\Omega/2(|M_j^{\pm}\rangle_{\mathcal{R}}{_{\mathcal{R}}\langle}S_j^{\pm}|+|S_j^{\pm}\rangle_{\mathcal{R}}{_{\mathcal{R}}\langle} M_j^{\pm}|)),\;
\end{array}
\end{equation}
where the energy $|V_c|$ includes possible off resonant energy shifts on atoms in singly occupied wells.  Including decay from the excited states at a rate $\gamma$, the non-unitary dynamics in the register is given by the non trace preserving master equation:
\begin{equation}
\dot{\rho}=-i/\hbar[H_I,\rho]-\gamma/2\sum_{j,\pm} (|M_j^{\pm}\rangle_{\mathcal{R}}{_{\mathcal{R}}\langle} M_j^{\pm}|\rho+\rho |M_j^{\pm}\rangle_{\mathcal{R}}{_{\mathcal{R}}\langle} M_j^{\pm}|).\;
\end{equation}
Assuming low saturation of the excited states, the dynamics in the ground state is
\begin{equation}
\begin{array}{lll}
\dot{\rho}_{S_{j}^{\pm},T}&=&-i\rho_{S_{j}^{\pm},T}(E(S_{j}^{\pm})+|V_{c}|)/\hbar+i(\rho_{T,T}-\rho_{S_{j}^{\pm},S_{j}^{\pm}})\sqrt{2} J/\hbar-\rho_{S_{j}^{\pm},T}\kappa\\
\dot{\rho}_{T,T}&=&i(\rho_{S_{j}^{\pm},T}-\rho_{T,S_{j}^{\pm}})\sqrt{2}J/\hbar\\
\dot{\rho}_{S_{j}^{\pm},S_{j}^{\pm}}&=&-i(\rho_{S_{j}^{\pm},T}-\rho_{T,S_{j}^{\pm}})\sqrt{2}J/\hbar-2\rho_{S_{j}^{\pm},S_{j}^{\pm}}\kappa .\;
\end{array}
\label{deneqs}
\end{equation}
\begin{figure}
\begin{center}
\includegraphics[width=1.0\linewidth]{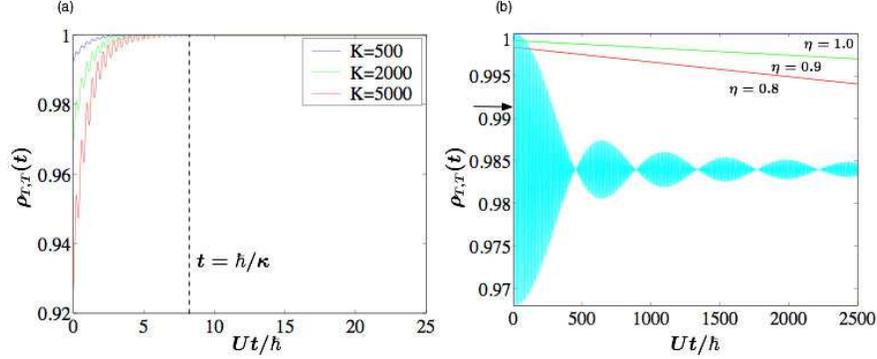}
\caption{\label{fig2}Population in the unit filled register state $|T\rangle_{\mathcal{R}}$ during continuous measurement of the register beginning in the Bose-Hubbard ground state $|\Psi_g\rangle$.  The plots show dynamics appropriate to tunneling in one dimension with $U/J=500$.  (a)  Quantum trajectories corresponding to a null measurement result for three different register sizes $K$.  The time scale to saturate the target state is independent of the number of qubits:  $t_{sat}\approx \kappa^{-1}$.  (b)  Long time dynamics for $K=501$, $N=551$ and finite detector efficiencies $\eta$.  The population in $|T\rangle_{\mathcal{R}}$ for $\eta=1$ is indistinguishable from one.  Also shown is the oscillatory dynamics at fundamental frequency $U$ described by Eq. \ref{fidelity} if the measurement is turned off after the target state is reached.  The arrow indicates $\rho_{T,T}(0)$.}
\end{center}
\end{figure}
Here the non-selective measurement induces phase damping of the coherences at a rate $\kappa=\Omega^2\gamma/(8 ((U/\hbar)^2+(\gamma/2)^2)$.  When the environment is monitored, for example by watching for photons, the population in the particle-hole states will change.  In the case of a null result, one obtains knowledge that the state is more likely in the target state and over time,  in the ``good" measurement regime  \cite{Milburn}, the probability to be in $|T\rangle_{\mathcal{R}}$ converges to one.  The ``good" measurement regime is given by the condition
$\Omega/\gamma\ll 1 <\hbar\kappa/2 \sqrt{K} J$,
which requires that the control field be strong enough to damp coherences on the time scale of the Bose-Hubbard dynamics but weak enough to not saturate the excited states.  Simulations of successful register preparation and maintenance by measurement are shown in Fig.~\ref{fig2}.
\subsection{Measurement at finite temperature}
Up to now we have been focusing on dynamics of pure states of the many body system.  
The selective measurement is an entropy decreasing map, because it damps amplitude in multiply
occupied wells.  Therefore it can be effective even for mixed states at finite temperature. 
We hereby assume a thermal distribution of  the eigenstates of the BH-Hamiltonian.  The overall effect of finite temperature is to increase
the weight of reduced Fock states other than the target state and
therefore to decrease the fidelity. 
For a temperature $T_d\approx \left( U-\epsilon_{(N-1)/2)} \right)/k_B $ 
states are populated which have more than one particle in one or 
more sites in the register.
The projective measurement is effective on these states and
the primary consequence of their presence is to reduce the initial fidelity of the system.
We can identify another temperature, $T_h\approx \left( \epsilon_{(N+1)/2}-\epsilon_{((K-1)/2)} \right)/k_B$ corresponding to the energy of the Fock state with a hole at site $(K-1)/2$. For $T\sim T_h$
there is appreciable population in eigenstates which have holes in the register.  The measurement is insensitive to population of states with holes
in the register, and the probability that the measurement does not project the system
into the target state, given a null measurement result, is then at least $1/e$ when the 
temperature is greater than $T_{h}$.
As a low initial fidelity can cause quantum jumps during the meaurement, 
large population of states with multiple occupancy
in the central sites should also be avoided. This leads to the rough
estimate $T_{max} \approx min\{T_d, T_h\}$.
\begin{figure}
\begin{center}
\includegraphics[width=0.45\linewidth]{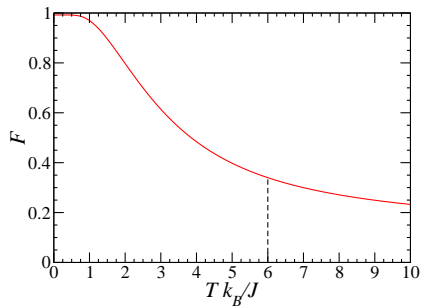}
\caption{\label{fidelity_m11}Equilibrium fidelity to be in the unit filled register state $|T\rangle_{\mathcal{R}}$ 
as a function of temperature 
in presence of the trap for $M=11$, $N=9$, $K=5$.  The relevant energies are
$U/J=60$ and $\delta/J=3.375$.  Here $T_d \approx 6.0 J/k_B$ and $T_h \approx 70 J/k_B$.  The dashed line indicates the scaled energy $T_dk_B/J$ where $F(T_{max})\approx 0.33988$. 
The fidelity at $T=0$ is $F(T=0)\approx 0.99208 $. }
\end{center}
\end{figure}
The exponential growth of the Hilbert space with particle number
makes the computation laborious.  We have obtained numerical results 
for a model system consisting of 
11 sites and 9 particles by exact diagonalization
of $H_{BH}$ in a Hilbert space of dimension $92378$.  The fidelity  is shown in 
Fig.~\ref{fidelity_m11}.  The coupling ratio is chosen to be 
$U/J = 60$ and the register
is defined by the central 5 sites.  In order to suppress tunneling of
holes into the register, the ratio $ \delta /J $ has been chosen to be 
very large, $ \delta /J \approx 3.375 $. 
This does not correspond to a typical experimental situation,
as it implies that the number of sites in the register for which
$\epsilon(j) < U$ is small (9, in this case).
The fidelity drops to values lower than $F<1/e$ at $T_d \approx 6.0 J/k_B 
= min\{T_d, T_h\}$, see caption of Fig.~\ref{fidelity_m11}. 
For different setups $T_d$ and $T_h$ can assume approximately the same value.
In the experimentally relevant setup discussed in relation to the
measurements, where $N=551$ and $K=501$, $T_d \approx 50J/k_B$, and
$T_h \approx 80J/k_B$.


\section{Summary and conclusions}
In summary, we have proposed a protocol to eliminate number
fluctuations in the many body ground state of a system undergoing 
Bose Hubbard dynamics.    For concreteness we have focused on preparation
of a register of cold atomic qubits in an optical lattice.  By considering the
free dynamics in both homogeneous and trapped systems we have
shown that there is a loss of fidelity to occupy the unit filled register that scales with the 
number of qubits because this 
is not a stationary state.  Because our analysis is based on first order
perturbation theory we established  the parameter regime where
our analytic approach is valid by considering second order
corrections.  The dynamical degradation of the fidelity once the register is
prepared is an issue that has to be considered not only
in our continuous measurement proposal but also in other
dissipative techniques proposed for unit filling initialization.  We have suggested one 
approach to correct for this by performing a 
continuous measurement on the system.  Numerical studies on small sized
systems indicate that this protocol can be made robust even for finite temperature systems which
will be important for real experimental implementations.

\section{Acknowledgements}
We thank Andrea Simoni and Zac Dutton for careful reading of the manuscript.  This work has received partial support through a grant from ARDA/NSA.

\end{document}